# Study of size and shell composition effect of TiO$_2$ core-shell mesoporous microsphere on UV absorption effectivity for photocatalytic application

**Yury E. Geints and Ekaterina K. Panina**
*V.E. Zuev Institute of Atmospheric Optics, 1 Acad. Zuev square, Tomsk, 634021, Russia;*
*Corresponding author e-mail: ygeints@iao.ru*

Abstract

Microdispersed photocatalysts based on titanium dioxide (TiO$_2$) in the form of hollow core-shell microspheres (microcapsules) with mesoporous structure are widely demanded in modern critical technologies related to the catalysis of various chemicals, solving environmental problems, and obtaining cheap fuel. To date, a number of experimental works are known, showing that geometrical parameters of microcapsules (size, shell thickness), as well as microstructural composition (nanosized metal additives, additional inner dielectric core- the "yolk") noticeably affect their photocatalytic activity. At the same time, a valuable physical description of the optical properties of porous microcapsules has not been presented in the literature so far. Using the finite element method, we perform a full-wave simulation of the optical field inside a hollow microsphere whose shell is assembled from multiple TiO$_2$ nanoparticles to form an irregular nanoporous structure. We give a physical explanation of the published experimental data on the optical activity of titanium-dioxide microcapsules and show that the existing theoretical models do not always give a correct interpretation of the observed empirical behaviors.

## 1. Introduction

The high level of industrialization in conditions of intensively expanding industrial production causes a number of serious problems associated, among others, with environmental pollution, which, in turn, leads to changes in climatic conditions. The development of technologies allowing effective air cleaning, purification of wastewater and various surfaces is currently an urgent task. One of the effective ways to solve this problem is the use of photocatalysts based on titanium dioxide (TiO$_2$). The process of photocatalysis is well studied and represents busting of chemical reactions under the action of light [1, 2]. TiO$_2$-based photocatalysts due to the environmentally friendly characteristics of the material, its nontoxicity combined with high photostability, and the ability to be easily synthesized at low temperatures, are widely demanded for hydrogen production and decomposition of pollutants [1, 3, 4]. In addition, titanium dioxide is a rather inexpensive material which also determines its demand in various production technologies.

To enhance the photocatalytic activity, $TiO_2$ modified nanostructures of various sizes and shapes, such as nanorods [5], nanofibers [6], nanotubes [7], etc. are used. In addition, the inclusion of other materials, such as noble metal particles, carbon-containing materials, or semiconductors in the structure allows in some cases significantly increasing the efficiency of the photocatalysis process [8, 9].

Among numerous structural solutions, core-shell composite structures are the most attractive from the point of view of technical realization and practical application. Multilayer objects assembled from materials with different properties can have a number of unique characteristics that determine different light absorption, chemical stability, and physical properties than their single-component counterparts. The most complete technology for obtaining multilayer photocatalysts is reviewed in detail in [10, 11].

The very idea of creating photocatalytic shell microstructures is based on the difference in the properties of materials forming a single structured composition thus improving its catalytic properties. In the case of a multilayer core-shell particle, the inner shell can act as an insulating layer eliminating the interaction between different materials. Depending on the practical application, not only the design but also the physicochemical properties of the core and shell vary. It is important to control both the thickness of the shell and its homogeneity and porosity. The latter factor is extremely important, since it is the porosity of the shell material that provides access of the reactant molecules to the encapsulated catalysts.

A systematic study of the effect of the size of mesoporous microcapsules assembled from crystalline titanium dioxide nanoparticles (NPs) (in anatase phase) on the photocatalysis (PC) efficiency of a dye (methylene blue) was undertaken in [11]. Experimental measurements showed that the geometrical parameters of the capsules, such as their size and thickness of the $TiO_2$ shell, significantly affect the PC activity of the suspensions. At the same time, there was a steady tendency for the decay rate of dye solution to decrease with increasing catalyst capsule size, which is directly related to the decrease in the total absorbed optical energy in the capsule shell. Additionally, the comparison of the photodissociation rate of dye solutions upon addition of titanium catalyst in the shape of ordinary nanoparticles (obtained by microcapsules crushing) and porous microcapsules showed a clear advantage of the latter (capsules) under the condition of equal volume weights of catalytic substances.

The enhancement of microcapsule PC activity by adding gold NPs is presented, e.g., in [12, 13]. In such a way, it was found in [12] that titanium dioxide microcapsules with diameters ranging from 200 nm to 1000 nm doped with several Au NPs in the core enhance hydrogen ($H_2$) yield during water catalysis. Interestingly, when the shell thickness of the capsules was relatively low, approximately from 65 to 140 nm (total capsule diameter 1is about 50-300 nm), the rate of the photocatalysis was higher than that with thicker capsule shells, from 190 to 390 nm (particle sizes 400-800 nm). Moreover, the very placement of the gold NPs in the microcapsule is also relevant to the PC efficiency. Among the two structural design options of metal-composite capsules, namely, on top of the capsule shell (Au-on-$TiO_2$) and inside the hollow core (Au@$TiO_2$), the latter structure shows the best result. Meanwhile, the greater the number of Au NPs in the core, the more efficient is the photocatalytic hydrogen evolution.

In Ref. [14], a sphere-in-sphere structure is proposed and it is shown that with an appropriate diameter of the inner sphere the optical radiation is concentrated in its volume more efficiently, which as claimed provides a better catalytic effect. According to [15], $SiO_2$-$TiO_2$ shell-yolk nanostructures enhance the photocatalytic release of $H_2$ due to multiple reflections of UV radiation between the boundaries of $TiO_2$ shell and $SiO_2$ core.

Two theoretical approaches have been proposed so far for the physical explanation the regularities observed in experiments related to the influence of the dimensional parameters of microcapsules on their PC activity. The first of them is given in the framework of the geometrical ray optics on the basis of the capsule model as a porous core-shell sphere, which provides multiple reflection and absorption of optical rays when they penetrate through mesoscale pores toward the particle core [10, 15, 16]. The second theoretical approach uses the analytical theory of optical wave scattering on a multilayer spherical particle with contrasting shells, this is known as the Lorenz-Mie theory [11, 17]. At the same time, it should be noting that both these strategies do not provide a complete understanding of the observed regularities and may even lead to paradoxical results, in particular, when interpreting the experimental data on the dependence of PC rate on the capsule size [11]. Moreover, the very applicability of ray optics is questionable for objects with mesowavelength dimensions ($R \sim \lambda$) and subwavelength internal structure (nanopores, crystal flakes). This motivated us to conduct a separate study of the optical activity of porous microcapsules by using the numerical simulation of light scattering by core-shell particles with complex microstructure.

In present work, we consider the problem of scattering and absorption of UV radiation by submicron bilayer particles of spherical shape, whose shell has an irregular nanorelief with multiple nanoscale pores formed by deposition of multiple titanium dioxide NPs. Using the finite element method (FEM), we numerically simulate the spatial distribution of the optical field in the vicinity of such core-shell microcapsules but with a different internal microstructure, which may include gold NPs (on top and inside the capsules), as well as an additional inner core - the "yolk". Mainly, our study is aimed on the evaluation the efficiency of optical energy absorption by the microcapsule shell of different structural design, which based on full-wave simulation will contribute to the development of a strategy to increase the PC activity of such microdispersed catalysts.

## 2. Optical model of porous core-shell microparticle and simulation details

Numerical simulations are performed using the COMSOL Multiphysics software package, which implements the solution of differential equations of wave optics by the finite element method (FEM). Structurally, a microcapsule is modelled as a multilayered microsphere with a non-absorbing aqueous core and a strongly light-absorbing solid-phase shell (Figs. 1a-c). Titanium dioxide is used as the shell material. To form a realistic solid shell of the microcapsule with chaotically arranged pores and caverns of nanometer sizes, we use an ensemble of spherical $TiO_2$ NPs the spatial position of which was set using an original software algorithm [18]. In this case, inside a given spherical layer with the thickness $D_{shell} = R_{cap} - R_{core}$ (see Fig. 1a) forming capsule shell, an array of NP center coordinates is programmatically generated obeying the random (normal) distribution with the possibility of partial overlapping (coalescence) of neighboring nanospheres. The diameter of $TiO_2$ NPs is fixed at 30 nm. This spherical shell generation procedure results in randomly arranged clusters of nanoparticles separated by irregular curvilinear voids.

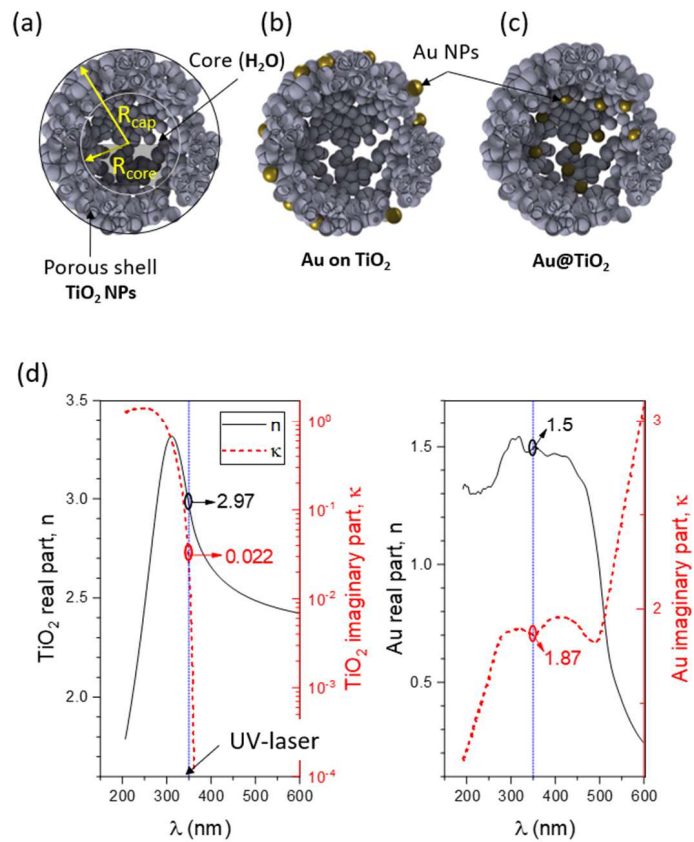

Fig. 1. (a-c) Structural types of composite mesoporous core-shell particles (microcapsules) used in the simulations of optical absorption. (d) Optical parameters of $TiO_2$ (anatase, crystalline film) and Au according to [19, 20].

If necessary, an array of Au nanospheres with random diameters distributed in the range from 15 nm to 40 nm can be additionally formed to add gold NPs to the capsule. The spatial position of Au NPs could be either outside the capsule shell (Fig. 1b, Au on $TiO_2$) or inside its core (Fig. 1c, Au@$TiO_2$). The spectral dispersion of the complex refractive index $m = n - j\cdot\kappa$ of titanium dioxide and gold NPs is shown in Fig. 1(d). Thus, for chosen UV laser wavelength $\lambda = 350$ nm delivering photons with sufficiently high energy to cover the $TiO_2$ bandgap of 3.2 eV by bound electrons [Gao2015], the calculation of the optical characteristics of the capsule shell gives the value, $m = 2.97 - j\cdot 0.022$. The microcapsule core and the surrounding medium are considered as water with refractive index $m_0 = 1.34$ and zero UV absorption.

### 3. Optical Absorption Characterization and Discussion
#### a. Impact of Au NPs

Consider the effect of gold NPs on the enhancement of PC production of composite microcapsules from the aspect of increasing their optical activity, i.e., a possible increase in light absorption by the particle shell. In this section, we discuss capsules of fixed size with $R_{cap} = 350$ nm and $R_{core} = 250$ nm only, which shell is assembled from dielectric nanoscale $TiO_2$ beads with volume fraction $\delta = V_{TiO_2}/V_{shell} = 0.65$, where $V_{TiO_2}$, $V_{shell} = 4\pi/3\left(R_{cap}^3 - R_{core}^3\right)$ are the total volume occupied by titanium dioxide and geometric volume of the shell, respectively.

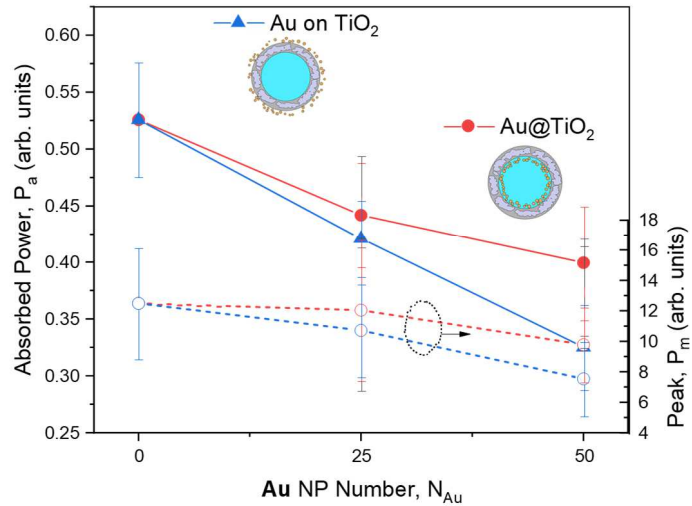

Fig. 2. Averaged $P_a$ and peak $P_m$ values of absorbed power in microcapsule shell in the dependence on Au NP number, $N_{Au}$. Error bars represent the STD values.

To quantify the effect of gold NPs on the optical absorption of titanium-dioxide mesoporous capsule, we perform a series of numerical calculations in Au@$TiO_2$ and Au-on-$TiO_2$ spatial configurations, and the results are shown in Fig. 2. Two quantities are plotted here, namely, the peak value $P_m$ of the absorbed optical power in the capsule shell and shell volume-averaged absorption $P_a$. Since the formation of both capsule shell and Au NPs ensemble is randomly generated each time the numerical model is loaded into the computer memory, each point in Fig. 2 is obtained by statistical averaging over 128 random realizations.

The analysis of the presented data shows that the addition of gold NPs in any region of the dielectric microcapsule has a negative effect in terms of the magnitude of the absorbed power by the particle shell. Meanwhile, Au NPs doped directly into capsule core reduce the optical absorption to a lesser extent than when Au beads are deposited outside the particle shell. This apparently counterintuitive conclusion, which does not coincide with the experimental data [12], nevertheless has an understandable physical explanation in terms of wave optics.

Indeed, consider Figs. 3(a-d) which show the distributions of normalized optical intensity $|\mathbf{E}|^2/E_0^2$ ($E_0 = 1$ V/m is incident electric field amplitude) and absorbed optical power $P_a$ in the principal cross-section of the microcapsule. Also, in these figures for the illustrative purpose, the streamlines of the Poynting vector $\mathbf{S} = (c/8\pi)\,\mathrm{Re}\left[\mathbf{E}\times\mathbf{H}^*\right]$ are plotted (**E**, **H** denote electric and magnetic field vectors, respectively, while $c$ is the speed of light) showing the direction of the optical energy fluxes inside and near the layered particle.

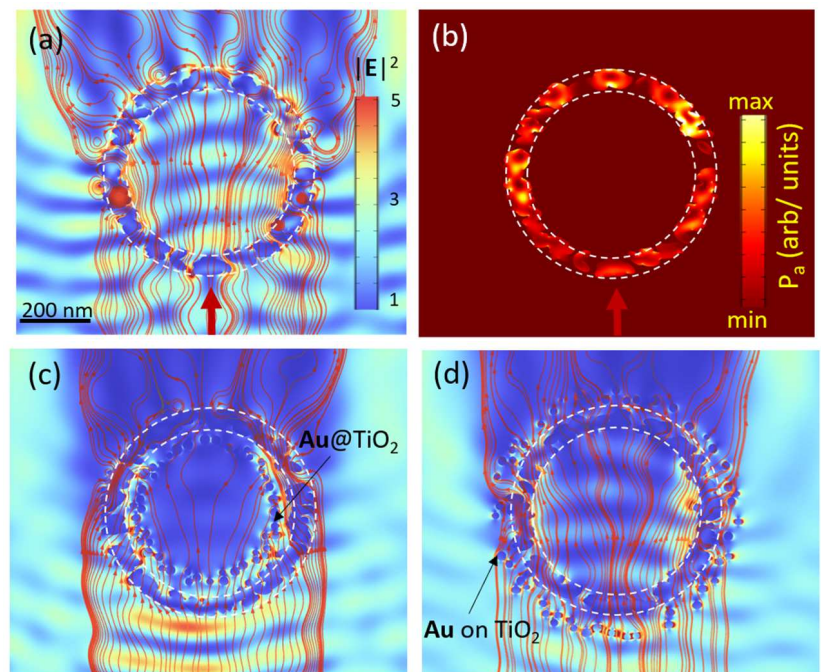

Fig. 3. (a, c, d) Optical intensity $|\mathbf{E}|^2$ and (b) averaged absorbed power density $P_a$ distributions in the principal cross-section of porous capsule without Au NPs (a, b) and in the cases of Au@TiO$_2$ (c) and Au-on-TiO$_2$ (d).

It is evident from these distributions that the scattering of the optical wave on a particle of mesoscale size ($R_{cap} \sim \lambda$) is accompanied by the appearance of multiple regions with microfocusing of laser radiation, where the optical intensity rises and the light absorption by the capsule shell increases (Fig. 3b). Basically, the optical focusing occurs inside the capsule shell, in the nanoscale gaps between neighboring clusters of NPs. As expected, in contrast to the theoretical considerations of Refs. [10, 15], no visible reflections of optical wave from the inner surface of the capsule shell could be observed when studying the full-wave optical field pattern in such small particles. Generally, the optical flux has a well-defined directionality, but nanopores inside the shell can influence the local density of the Poynting vector **S** dividing it into small-scale fluxes and bending them at different angles with respect to the direction of the energy mainstream.

The addition of gold NPs outside or inside the microcapsule noticeably rearranges the structure of optical fluxes. When Au NPs are inside the microparticle shell (Fig. 3c), the whole ensemble of metallic NPs acts as a diffraction grating with a deep subwavelength spatial period which partially blocks electromagnetic wave propagation into the shadow regions of the capsule. In this sense, the array of Au NPs resembles the well-known Faraday cage, which prevents the electric field from passing into the shielded area (capsule core). As can be clearly seen in Fig. 3c, the Poynting streamlines bend near the illuminated part of the microparticle indicating the tendency of the optical flux to go around the spherical metal nanograting. In this case, the energy flux is most concentrated inside the side surfaces of the capsule shell as well as in the gap between the shell and the layer with gold NPs, while the back of the particle falls into the region of deep shadow.

If Au NPs are located outside the capsule shell, as shown in Fig. 3d, the Au "Faraday cage" surrounds the whole particle and effectively blocks the optical radiation. However, for the same number of gold NPs ($N_{Au}$), Au-on-TiO$_2$ structure demonstrates more sparse gold grating than when Au NPs are deposited inside the capsule core, thus allowing optical radiation for partially reaching the absorbing shell and participate in the photocatalytic generation of electron-hole pairs.

Thus, the concept of increasing the optical absorption by the porous capsule due to the addition of gold NPs is not confirmed within the framework of our full-wave modeling. On the contrary, a slight decrease in the absorption efficiency of the shell is observed, which becomes more noticeable with increasing Au NP number. Obviously, the actual role of the metallic nano-additive is not in mediating the optical properties of the parent capsule, but in positively changing the free-carrier dynamics arising at the contact sites with titanium dioxide during its absorption of optical photons. In this case, as shown in numerous studies [12, 21, 22], nanoparticles of some metals (Au, Pt, Pd) increase the photocatalytic activity of composite microcapsules due to the suppression of unwanted electron-hole recombination on the surface of TiO$_2$. A model explaining this physical process was proposed in [12]. In contact with a metal, the photoinduced electrons in titanium dioxide molecules which moved from the valence band to the conduction band under the influence of photons, migrate into metal molecules due to Fermi levels bending. This improves the stability of positively charged holes appearing on the surface of $TiO_2$ and increases the lifetime of free charges, which, in turn, increases the rate of PC reactions. Interestingly, in a recent study [23] it is shown that the rate of catalase-like enzyme reaction in hydrogen peroxide ($H_2O_2$) also reduces when such a micro dispersed catalyst in the form of polymeric microparticles with encapsulated in the core gold NPs is used.

### b. Capsule Size Effect

At the following phase of simulation, we studied the size effect of the microcapsule on the value of its optical absorption. To this end, two series of simulations are performed. First, the size of the inner core $R_{core}$ of bilayer particle is varied, while the thickness of its shell $D_{shell}$ is fixed. This is equivalent to changing the geometric midsection of the scatterer while maintaining its optical thickness. In contrast, in the second series of simulations, it is assumed a constant core radius but the shell thickness is allowed to vary, which changed both the optical wave propagation distance inside the absorbing nanocomposite medium and the extinction cross-section of the whole particle. Importantly, the volume fraction of $TiO_2$ in the shell is also fixed, $\delta = 0.5$. The statistically averaged values of the volume-averaged and peak absorption power are shown in Figs. 4(a) and (b).

Generally, the highest optical absorption exhibits the capsules with sizes on the order of the wavelength of incident optical radiation, $R_{cap} \sim \lambda$. In the case of a fixed shell thickness (Fig. 4a), increasing the radius of the composite microparticle leads to a near monotonic drop in the volume-averaged absorbed optical power $P_a$, which was previously observed experimentally in [11]. At the same time, the peak optical absorption density $P_m$ that is achieved in the $TiO_2$ shell varies rather weakly around some constant level, $P_m \approx 12$ (in arbitrary units). Note that here and throughout the paper, for convenience, we give the absorbed power in some arbitrary units, since it is important to know not the absolute value of the absorption, but to understand the trend of its variation when measuring the parameters of the photonic structure (capsule). Importantly, when the capsule shell is homogeneous, i.e., without pores and voids (dashed curve in Fig. 4a), the optical absorption power in such bilayer particle is always lower than for the case of an inhomogeneous porous shell. In the mesowavelength range of capsule radii, $R_{cap} \sim 200$ to $300$ nm, this difference can be up to 1.5 times.

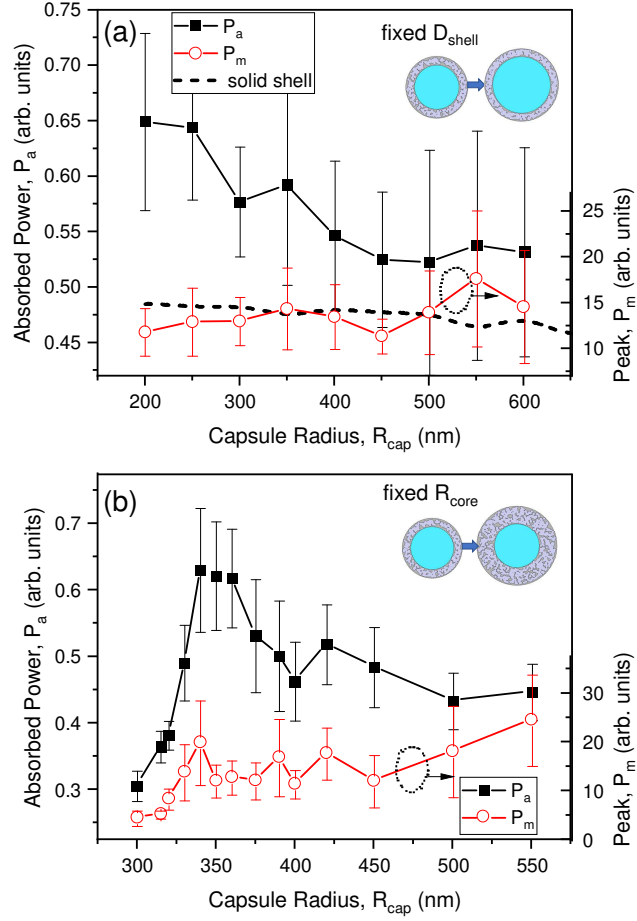

Fig. 4. Effect of (a) core size and (b) microcapsule shell thickness on optical absorption. Simulations were performed at a constant $TiO_2$ fraction $\delta = 0.5$ and either (a) fixed shell thickness $D_{shell} = 200$ nm, or (b) constant core radius, $R_{core} = 250$ nm. $P_a$ values for a capsule with a solid homogeneous $TiO_2$ shell are shown in (a) by the dashed curve.

The effect of increasing the size of the capsule by increasing its shell thickness while keeping the size of the liquid core fixed is illustrated in Fig. 4(b). Clearly, when the absorbing shell becomes too thin, i.e., such that dielectric $TiO_2$ NPs (with a diameter of 30 nm) can form only a single layer during its deposition into the shell, the averaged absorption power $P_a$ of the capsule drops sharply. This is because the optical wave passes through the thin polydisperse shell practically without scattering on the islands of titanium dioxide, which is confirmed by the reduced values of peak absorption in the shell $P_m$. On the contrary, with a more thick shell, $D_{shell} > 200$ nm ($R_{cap} > 450$ nm), the scattering of radiation by the capsule begins to prevail over absorption in the shell and the effective capsule absorption decreases despite the fact that the optical field maxima become sharper ($P_m$ values increase). The result of the competition between these oppositely directed tendencies is the absorption maximum observed in capsules with $R_{cap} \sim 320$ to 370 nm.

It is worth noting that, in general, the behavior of the absorbed power at the capsule size resembles the well-known dependence for optical extinction efficiency $Q_{ext}$ of a dielectric sphere by varying its size as predicted by the Lorenz-Mie theory [24]. This dependence is characterized by the presence of an absolute maximum at $R \sim \lambda$ and an oscillating tail in the region of large sphere sizes having an asymptote at a value of $Q_{ext} = 2$.

### c. Effect of Shell Porosity

The question of the influence of polydisperse shell density on its absorption activity is also important. To this end, the simulations of absorption by a core-shell particle with fixed dimensional parameters are carried out while varying the fraction $\delta$ of titanium dioxide NPs in the shell. The results of calculations of the parameters $P_a$ and $P_m$ are presented in Fig. 5.

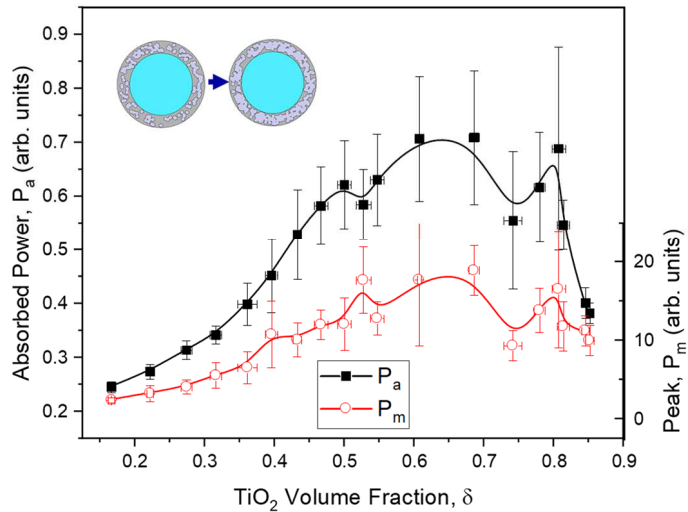

Fig. 5. Effect of capsule shell porosity. Absorption by capsule with $R_{core} = 250$ nm and $R_{cap} = 350$ nm when changing the fraction $\delta$ of $TiO_2$ NPs in shell. Solid lines are bold spline fitting.

As expected, we observe a monotonic growth of absorption with increasing shell density, which is explained both by increasing the mass fraction of the light absorbing component and reducing the gaps (pores) between clustered titanium dioxide NPs. The growth of optical absorption by the core-shell sphere with $\delta$ saturates when the $TiO_2$ volume fraction reaches the value $\delta \approx 0.5$. When the shell density is further increased, averaged absorption $P_a$ starts to drop to the value of about $P_a \approx 0.48$ which equals to a capsule with a solid homogeneous $TiO_2$ shell. This elucidates the importance of the presence of gaps between the NP clusters in shell, i.e., the presence of nanopores, in view of the realization of the anomalous optical field intensity enhancement in these nanovoids. When $TiO_2$ volume fraction approaches a unity ($\delta \rightarrow 1$) and the pores in the capsule shell disappear, the maximal and averaged values of capsule absorption also decrease.

### d. Influence of Additional "Yolk"

One of the possible designs for a photocatalyst based on core-shell microparticles is a structure in which an additional scattering core is added in the capsule, referred to in the literature as a "yolk" [10, 15]. As believed, the role of this yolk is to add an extra reflection to the optical rays that have entered the capsule through the porous shell. As a consequence, the optical path of rays inside the particle becomes longer and the absorption of light by the shell increases.

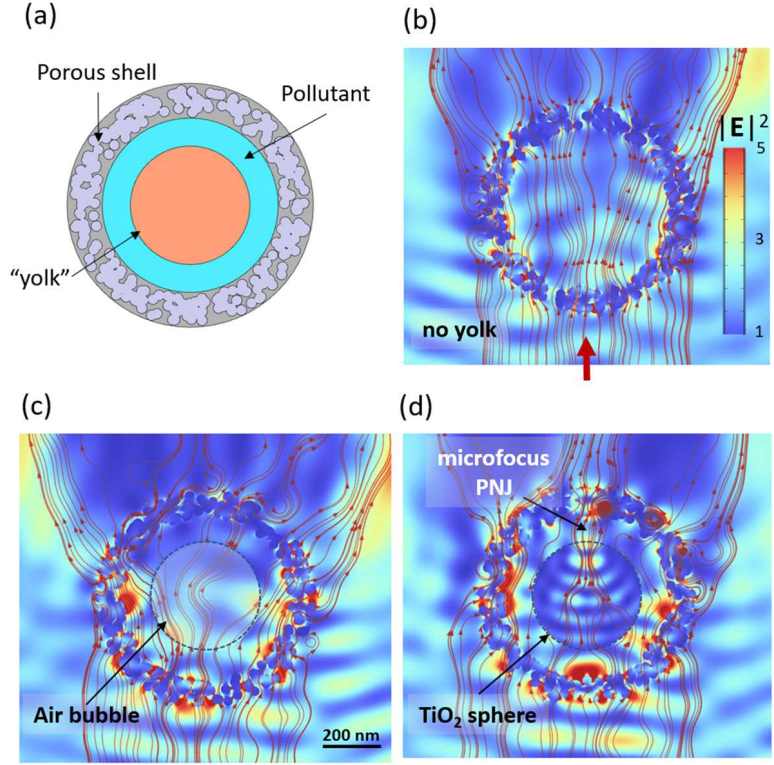

Fig. 6. (a) Schematics of a core-shell particle with an additional inner core (yolk). (b-d) Structure of optical intensity $|\mathbf{E}|^2$ and energy fluxes (Poynting vector $\mathbf{S}$) in capsules (b) without yolk and with yolk in the form of (c) air bubble with $m_y = 1$ and (d) TiO$_2$ microsphere with $m_y = 2.97 - j \cdot 0.022$.

We numerically simulate such a photonic structure by adding a spherical core with different refractive index $m_y$ inside the mesoporous capsule. This can be a 200 nm (in diameter) TiO$_2$ microsphere, i.e., a particle from the same material as the shell, or an air bubble with refractive index equal to one. The simulation results are shown in Figs. 6(b-d) as the cross-sectional distributions of the normalized optical intensity $|\mathbf{E}|^2$ and the streamlines of the Poynting vector $\mathbf{S}$. For comparison, Fig. 6(b) shows the situation with a core-shell microparticle without an inner yolk.

As seen, depending on the magnitude of the optical contrast of the inner liquid core ($m_0 = 1.34$) and the yolk ($m_y$), the optical energy fluxes rearrange their structure in different ways. Indeed, in the case of the air bubble in Fig. 6(c), after passing the particle shell and hitting the yolk, the optical wave experiences scattering on an optically low-density medium. Because the optical contrast "yolk-core", ($m_y$-$m_0$), is negative, this leads to an additional angular divergence of the energy fluxes passing through the center of the capsule (cf. with the case of the pure core-shell particle in Fig. 6b). As a result of this defocusing, the optical fluxes more intensive illuminate the lateral regions of the shell and can provide a modest increase in absorption.

In the case of the titanium dioxide core in Fig. 6(d), on the contrary, the denser core provides focusing of the central radiation beam with the formation of a narrow nanoscale light beam, called the photonic nanojet (PNJ), near the shadow surface of the yolk [25, 26]. This PNJ is projected onto the inner shell of the capsule and also leads to an additional increase in light absorption.

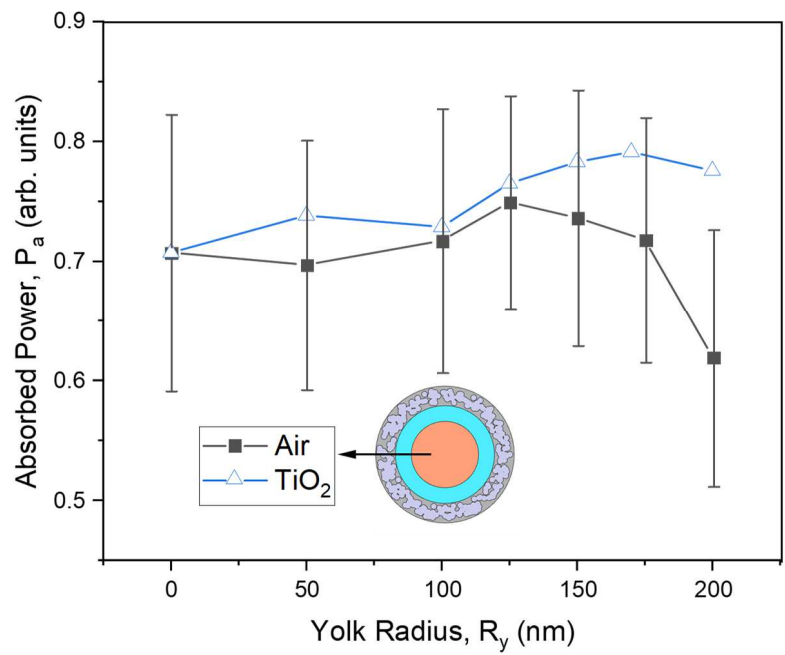

Fig. 7. "Yolk" effect on optical absorption by a capsule. Averaged optical absorption of core-shell-yolk particle with $R_{cap}$ = 350 nm and $D_{shell}$ = 100 nm at different parameters of the inner yolk at $\delta$ = 0.6.

Quantitatively, the effect of the additional inner core on the absorption efficiency of the capsule is shown in Fig. 7. During the simulation, we change the radius $R_y$ and the complex refractive index of the yolk while maintaining the volume fraction of $TiO_2$ NPs in the particle shell. One can see a general tendency for a moderate (~ 15%) increase in the averaged absorption by the capsule shell when $TiO_2$ yolk with a radius of $R_y$ = 100-170 nm is added. A larger yolk size becomes no longer effective as the additional core starts to approach the capsule shell, $R_y \to R_{core}$, which increases the scattering cross-section of the particle but decreases its absorption. In addition, the fraction of liquid analyte inside the hollow sphere to realize photocatalytic decomposition is reduced, which also has a negative effect on the photocatalytic activity. If one studies the change in hollow particle absorption when adding yolk with the negative contrast resulting from the inclusion, e.g., of air bubble, it is also possible to increase the absorption of the capsule shell. However, this increase is lower than that from $TiO_2$ yolk and the drop in $P_a$ values at the larger size of the additional core is more pronounced.

## 4. Conclusion

To conclude, we theoretically consider the problem of increasing the optical absorption of spherical core-shell microparticles (microcapsules) formed during clustering of titanium dioxide NPs for the purpose of enhancing their activity in the photocatalysis processes. On the basis of a full-wave numerical model of a hollow microsphere with a mesoporous polydisperse shell, the numerical calculations of UV absorption ($\lambda = 350$ nm) are carried out when the structural parameters of the capsule are changed, including the addition of Au NPs inside and outside capsule shell as well as adding an extra core (yolk). We show that in the UV range the best absorption efficacy possess the hollow particles with a radius of the order of the illuminating wavelength, i.e. with $R_{cap} \sim 300\text{-}400$ nm, and a porous shell thickness between 70 nm and 150 nm. Our findings are in good agreement with the experimental work [11]. Additionally, the importance of the porosity of the microcapsule shell, i.e., the presence of gaps between the $TiO_2$ NP clusters constituting the shell, is clearly demonstrated. It turns out that the nanopores contribute to the huge enhancement in the optical intensity realizing within these nanovoids and thus increase the optical absorption of the whole particle.

Surprisingly, the concept on the role of gold NPs or other metals in enhancing the optical activity of hollow microparticles used as a titanium-dioxide catalyst is valid only to some extent. Indeed, a thorough analysis of the optical energy fluxes inside the hollow porous microsphere with Au NPs additive shows that the presence of gold NPs outside or inside the microcapsule noticeably rearranges the structure of the energy flux. However, the overall optical absorption by the shell is rather negatively affected, which is amplified when the number of gold NPs increases. We believe that the positive role of the metal nano-additive established in previous studies [12, 21, 22] is in changing the dynamics of free carriers arising at the contact sites with titanium dioxide atom when it absorbs UV photons.

Finally, we study the effect on capsule optical absorption of an extra dielectric core – the "yolk", which as suggested earlier [15], should increase the optical path of light rays inside the hollow particle due to additional reflections. In our simulation, the effect of increasing the total absorption by the microcapsule with the yolk core is indeed established both with a dense $TiO_2$ yolk and when an air bubble is placed inside the capsule. However, it is worthwhile noting that this increase in overall absorption is low, no more than 15%, and is observed mainly for $TiO_2$ extra nucleus with a radius from 100 to 170 nm.

**Funding**. Ministry of Science and Higher Education of the Russian Federation (IAO SB RAS).

**Disclosures**. The author declares no conflicts of interest.

**Data availability**. Data underlying the results presented in this paper may be obtained from the authors upon reasonable request.